# DETECTION OF PERFUSION ROI AS A QUALITY CONTROL IN PERFUSION ANALYSIS

**Alkhimova S.M.**  PhD

National Technical University of Ukraine "Igor Sikorsky Kyiv Polytechnic Institute"

*Abstract. In perfusion analysis automated approaches for image processing is preferable due to reduce time-consuming tasks for radiologists. Assessment of perfusion results quality is important step in development of algorithms for automated processing. One of them is an assessment of perfusion maps quality based on detection of perfusion ROI.*

**PURPOSE**

The non-invasive methods of perfusion evaluating play important role in the diagnostic of cerebrovascular and oncological diseases. On current moment the perfusion analysis based on dynamic susceptibility contrast (DSC) magnetic resonance (MR) exams is widely used for these purposes. In this technique, the passage of contrast agent particles through the volume causes signal changes within tissues. This effect is monitored by a dynamic series of exam images. Image processing on a pixel-by-pixel basis provides approximate quantification of different hemodynamic parameters and uses to generate color-coded perfusion maps for them.

All methods of image processing that target to calculate the perfusion parameters are based on changes in the concentration of contrast medium during its passage through the vascular system. By the way the grounds of such methods have changed with the evolution of computer technology. Modern software for DSC perfusion analysis in most cases uses different variations of deconvolution method (model-dependent, model-independent, statistical approaches). In this method quantification of blood flow is done by deconvolution of the tissue time-concentration curve with arterial time-concentration curve (arterial input function). And, as a result, the method gives approximate values for such perfusion parameters as blood flow (or perfusion), blood volume and mean transit time. It should be mentioned that hemodynamic parameters are calculated with fitting of a gamma-variate function to the first pass of the time-concentration curve before deconvolution applying. Fitting-based perfusion analysis is time consuming and, even more important, the corresponding results are high sensitive to the image noise. Due to the fast and simplicity of calculations, the summary perfusion parameters, which are based on the original values of the time-concentration curves, remain widely used as separate method for perfusion analysis as well as partially involved in overall deconvolution based perfusion report.

Nowadays, availability of a large number of software variants that provide different implementations of the algorithms for perfusion analysis cause significant differences between the results of perfusion maps visualization and quantitative values. Accuracy and reliability assessment of implementation of the algorithm for perfusion analysis can be provided through using digital phantom data [1, 2]. Such assessment is based on visual and quantitative results correlation with true values that were generated using a tracer kinetic theory. However, perfusion analysis results on clinical data can vary a lot even within the same software usage on the same dataset. This is explained through the different detection of perfusion region of interest ROI [3, 4].

### MATERIALS AND METHODS

To assess perfusion maps quality through the different detection of perfusion ROI DSC head MR images were acquired on a 3.0 T clinical scanner (Achieva, Philips Healthcare, Best, the Netherlands) from 6 patients with cerebrovascular disease. Scan parameters were: repetition time = 1500 ms, echo time = 30 ms, flip angle = 90º, field of view = 23 x 23 cm, image matrix = 256 x 256, slice thickness = 5 mm, and gap = 1 mm. 17 slices were scanned with 40 dynamic images for each slice. Contrast medium (Gadovist, Bayer Schering Pharma AG, Berlin, Germany) with a dosage of 0.1 mmol/kg body weight was injected at a rate of 5 mL/sec, followed by a 10-mL bolus of normal saline also at 5 mL/sec. All images were collected in 12-bit DICOM (Digital Imaging and Communication in Medicine) format.

Widely used thresholding technique was analyzed as automated perfusion ROI detection. Thresholding was done with different threshold values from analyzed image intensities window for both only low and low & high intensity pixels extraction. The results of automated perfusion ROI detection were compared with a reference standard (manually marked ROI of the brain perfusion data by an experienced radiologist and confirmed by a second radiologist).

### RESULTS

The figure shows original DSC head image and visualization results of mean transit time (MTT) perfusion map when perfusion ROIs were detected in different ways: automated detection of perfusion ROI with only low intensity pixels extraction, automated detection of perfusion ROI with low & high intensity pixels extraction, manually marked perfusion ROI.

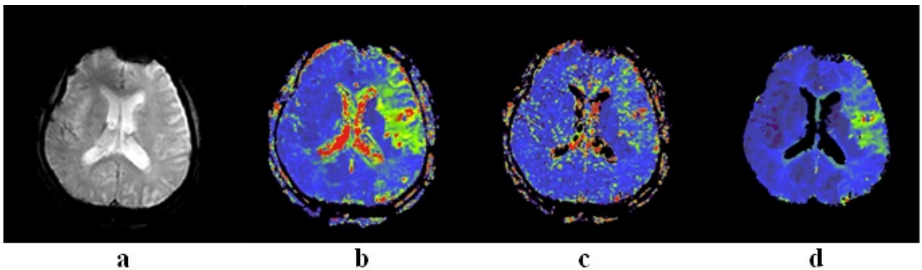

a   b   c   d

**Figure: (a)** Sample of DSC head image. **(b)** MTT map when perfusion ROI was automatically detected with only low intensity pixels extraction. **(c)** MTT map when perfusion ROI was automatically detected with low & high intensity pixels extraction. **(d)** MTT map when perfusion ROI was manually marked.

Sample DSC head image and threshold values for automated perfusion ROI detection were selected for present results with average similarity, i.e. MTT maps visual results with average correlation (correlation coefficient, r = 0.54) between automated and manual ROIs detection among all analyzed samples. MTT map is visualized with the same LUT scheme and within the same intensities window.

**CONCLUSION**

In conclusion, poor automated detection of perfusion ROI can lead to degradation of perfusion maps quality. Obtained results show weak similarity between perfusion maps with automated detection of perfusion ROIs compare to the manual ones. Perfusion ROI detection as well as LUT scheme usage and display values range of result maps have to be standardized quality control in perfusion analysis.